 \documentclass[pra,showpacs]{revtex4}
 \usepackage{amssymb} \usepackage{graphicx} \graphicspath{ {images/} }
  \usepackage{dcolumn} \usepackage{amsmath}
 \usepackage{tikz}
  \usetikzlibrary{shapes,backgrounds}
\usepackage{color}

\begin{document}
\title{Twistors from Killing Spinors alias Radiation from Pair Annihilation I: Theoretical Considerations}
 \author{\"{O}zg\"{u}r A\c{c}{\i}k}
\email{ozacik@science.ankara.edu.tr}
\address{Department of Physics,
Ankara University, Faculty of Sciences, 06100, Tando\u gan-Ankara,
Turkey\\}

\begin{abstract}
This paper is intended to be a further step through our Killing spinor programme started with Class. Quantum Grav. \textbf{32}, 175007 (2015), and we will advance our programme in accordance with the road map recently given in arXiv:1611.04424v2. In the latter reference many open problems were declared, one of which contained the uncovered relations between specific spinors in spacetime represented by an arrow diagram built upon them. This work deals with one of the arrows with almost all of its details and ends up with an important physical interpretation of this setup in terms of the quantum electrodynamical pair annihilation process. This method will shed light on the classification of pseudo-Riemannian manifolds admitting twistors in connection with the classification problem related to Killing spinors. Many physical interpretations are given during the text some of which include dynamics of brane immersions, quantum field theoretical considerations and black hole evaporation.
\end{abstract}

\maketitle

\section{Introduction}
 In the above mentioned arrow diagram there were two unanalyzed nontrivial arrows, the brown one illustrating the possibility for constructing Dirac spinors from parallel spinors in curved spacetimes and the blue one representing the production of twistors from Killing spinors. As told in the abstract, the blue arrow has a physical interpretation in terms of the quantum electrodynamical pair annihilation process, where Killing spinors correspond to the electron-positron pair and twistors correspond to the pure radiation field. The present interaction for some of its reasonable features is furthermore thought to be gravitational, but it stays as a conjecture that the resulting twistors should be related to the graviton or to gravitino. A method in this direction seems to exist, namely the spin raising/lowering of non-gravitational fields in curved spacetimes \cite{Penrose Rindler, Benn Kress spin, Charlton}. In the light of some invaluable and interesting results obtained in \cite{Acik III}, the proof of the above conjecture is hoped to be improved pretty much in a future work. There the full power of geometric Killing spinors will be given, relating them to almost all important equations of mathematical physics \cite{Acik 2017}.\\

The method of the present paper, to our guess,  will also shed light on the classification of pseudo-Riemannian manifolds admitting twistors \cite{Baum Leitner,Lichewski} in connection with the classification problem related to Killing spinors \cite{Kath}.\\

We prefer to use coordinate-free differential geometry for the sake of notational elegance and brevity and also for clarity in direct geometrical or physical interpretations. Unfortunately, the usage of componentwise tensor calculus is more common in physics community; closing this gap could only be possible by transforming one language to the other. We tried to do this language translation at least for a set of equations that we feel important and any reader can use this example as a dictionary in between; this is done in Appendix C. Pre-metric notions are handled through the usage of ''Cartan calculus on manifolds'' \cite{Tucker Rutherford, Burton}, the participation of a metric tensor field into the set of spacetime structures \footnote{Of course in almost all geometric theories of gravity metric is the fundamental field, especially in General Relativity.} forces us to use the full power of ''Clifford bundle formulation of physics and geometry''. This latter choice is precious for admitting a local equivalence between spinor fields and the sections of minimal left ideal bundle of the Clifford bundle. A good account of our tradition can be found in \cite{Benn Tucker, Tucker daktilo}. Throughout the text we assume a Riemannian spacetime \footnote{ Although we have used the term pseudo-Riemannain in the abstract and in the introduction we go against this usage with the same lines of thought as R. Toretti \cite{Toretti}, and instead we prefer to use Riemannian spacetime generally and especially Lorentzian spacetime for manifolds endowed with one time-like direction in a local $g$-frame.} i.e. a smooth manifold endowed with an indefinite metric $g$ and its Levi-Civita connection $\nabla$.\\

The organisation of the paper is as follows. In section II an explicit method for constructing conformal Killing-Yano (CKY) forms from Killing-Yano (KY) forms and closed conformal Killing-Yano (CCKY) forms is given; before that a new form for the definitive equation for KY forms is built up. Then the first order symmetry operators for massive and massless Dirac equations are reconsidered and are related in an elegant way in the view of the above setup. An intuitional question, which leads to the core of this paper, is asked at the end of section II. The positive answer of this main question is given in terms of a proposition and its proof. The analysis is bifurcated into fermionic and bosonic sectors both of which contain geometric identities that are derived in the corresponding subsections; the fermionic sector is detailed by considerations on charge conjugation, time reversal, helicity and their relation to Killing reversal due to the preceding main physical interpretation. After the conclusion of our paper at section IV, four appendices are given at the end.

Appendix A is based on the calculation of the adjoints of the symmetry operators associated to KY and CCKY forms and ends up with a comment on symmetry algebra. Appendix C contains the derivation of the coordinate expressions of our primitive set of equations which we feel important for the understanding of general reader who are used to the notation of tensor components. Appendix D gives a derivation of equations (31) and (32).

%***********************************************************************

%************************************************************************
\def\nblx#1{\nabla_{X_{#1}}}
\def\sbl#1{(\psi\overline{\psi})_{#1}}

\section{Building Up Conformal Killing-Yano Forms}

\subsection{A different perspective for defining Killing-Yano Forms}
The important roles played by KY forms and CKY forms are various. They define the hidden symmetries of the ambient spacetime generalising to higher degree forms the ones induced by Killing and Conformal Killing vector fields, and they respectively can be used for setting up first order symmetry operators for massive and massless Dirac equations in curved spacetimes \cite{Benn Kress2, Acik Ertem Onder Vercin1}. These symmetry operators reduce to the Lie derivative on spinor fields for the lowest possible degree, so if carefully worked out they may define Lie multi-flows of spinor fields in a classical spacetime. Achieving this shall also extend the concept of Lie multi-flows to form fields and multi-vector fields. As a consequence, the analysis of higher rank infinitesimal symmetries will heavily simplify \cite{Ertem 2016a, Ertem 2016c}. On the other side, via Noether's theorem, they induce conserved quantities associated to geodesics of (point) particles \cite{Hughston Penrose Sommers and Walker} and more generally to world-immersions of $p$-branes (\cite{Kastor Trachen, Acik Ertem Onder Vercin2, Acik Ertem 2016}; where there is much work to do for the later case \cite{Acik JMP 2016}. Also the definition of some charges for higher dimensional black holes rely on the existence of these symmetry generating totally anti-symmetric tensor fields \cite{Krtous et al}. \\

Now we want to redefine KY forms from a different perspective that is more general. The usual definition of a KY-form of degree $p$ is the solution of
\begin{eqnarray}
\nabla_X\omega=\frac{1}{p+1}i_{X}d\omega.
\end{eqnarray}
Exterior derivative $d$ and co-derivative $d^{\dag}=*^{-1}d*\eta$ can be given by the Riemannian relations $d=e^a \wedge \nabla_{X_a}$ , $d^{\dag}=-i_{X^a}\nabla_{X_a}$. If $T$ is a mixed tensor field of covariant degree $r$ and contravariant degree $s$, its covariant differential $\nabla T$ which has one more contravariant degree than $T$ is defined as
\begin{eqnarray}
(\nabla T) (X,Y_{1},...,Y_{r},\beta^{1},...,\beta^{s})=(\nabla_{X} T) (Y_{1},...,Y_{r},\beta^{1},...,\beta^{s})
\end{eqnarray}
in terms of vector fields $X,Y_{i}$ and co-vector fields $\beta^{j}$. Also remembering the identity
\begin{eqnarray}
(i_{X}\phi)(Y_{1},...,Y_{p-1})=p \phi(X,Y_{1},...,Y_{p-1})
\end{eqnarray}
for any p-form $\phi$, we can define KY-forms in a different manner. If we write (1) as
\begin{eqnarray}
(\nabla_{X}\omega)(Y_{1},...,Y_{p})=(\frac{i_{X}d\omega}{p+1})(Y_{1},...,Y_{p}) \nonumber
\end{eqnarray}
then by using (2) and (3) for the left and right hand sides respectively we obtain
\begin{eqnarray}
(\nabla\omega)(X,Y_{1},...,Y_{p})=(d\omega)(X,Y_{1},...,Y_{p})
\end{eqnarray}
which is valid for any $X,Y_{i}\in \Gamma TM$. As a result the new defining equation for a KY form is
\begin{eqnarray}
(\nabla-d)\omega=0,
\end{eqnarray}
and is not necessarily homogeneous. An integrability condition equivalent to Poincar\'{e}'s lemma is $\nabla^{2}\omega=0$. Remembering the definition of the alternating idempotent map; i.e.
\begin{eqnarray}
 \nonumber % Remove numbering (before each equation)
  Alt:\Gamma T^{r}(M)=& \longrightarrow& \Gamma \Lambda^{r}(M) \\ \nonumber
  G\quad =& \rightarrowtail& Alt(G
\end{eqnarray}
such that $$Alt(G)(Y_{1}, Y_{2},..., Y_{r})=\frac{1}{r!}\sum_{\sigma \in S(r)}sign(\sigma)G(Y_{\sigma(1)}, Y_{\sigma(2)},..., Y_{\sigma(r)}),$$
it is possible to write (5) as $$\Big((1-Alt)\nabla\Big)\omega=0.$$

\subsection{Adding Closed Conformal Killing-Yano forms to Killing-Yano forms}
In this section our main objects will be CKY forms and we will give a trivial method for their construction. CKY form equation is
\begin{equation}
\nabla_X\rho=\frac{1}{p+1}i_{X}d\rho-\frac{1}{n-p+1}\widetilde{X}\wedge d^{\dag}\rho
\end{equation}
 where an inhomogeneous generalisation was given in \cite{Acik Ertem 2015}. We want to emphasize the duality between the space of Killing-Yano forms and the space of CCKY forms, that is if $\widehat{\omega}$ is a CCKY $p$-form then its Hodge dual $*\widehat{\omega}$ is a KY $(n-p)$-form and vice-versa. Although the usual definition of a CCKY $p$-form $\widehat{\omega}$ is given by
\begin{equation}
\nabla_X\widehat{\omega}=-\frac{1}{n-p+1}\widetilde{X}\wedge d^{\dag}\widehat{\omega},
\end{equation}
it is trivial from (6) that this equation can also be written as
\begin{equation}
(\nabla^{\dag}-d^{\dag})\widehat{\omega}=0
\end{equation}
where $\nabla^{\dag}:=*^{-1}\nabla*\eta$. $\widetilde{X}$ is the $g$-dual of the vector field $X$ and $\eta$ is the main automorphism of the tensor algebra; additionally by analogy to Hodge-de Rham operator $\displaystyle{\not}d:=d-d^{\dag}$ we introduce $\displaystyle{\not}\nabla:=\nabla-\nabla^{\dag}$ which should not be confused by the Dirac operator $\displaystyle{\not}D$; also $\nabla^{\dag\dag}=(-1)^n \nabla$. In a Riemannian spacetime, the $\rho$ in (6) satisfies (7)/(1) if it is closed/co-closed; so it is trivial that if $\omega$ is a KY $p$-form and $\widehat{\omega}$ is a CCKY $p$-form then we can built up a CKY $p$-form simply as $$\rho=\omega+\widehat{\omega}.$$
\textbf{Proof:} Covariant derivative of $\rho$ with respect to a vector field $X$ is $$\nabla_{X}\rho=\nabla_{X}\omega+\nabla_{X}\widehat{\omega},$$ using (1) for the first term and (9) for the second term at the right hand side gives $$\nabla_{X}\rho=\frac{1}{p+1}i_{X}d\omega-\frac{1}{n-p+1}\widetilde{X}\wedge d^{\dag}\widehat{\omega}.$$ Adding $\widehat{\omega}$ to $\omega$ and $\omega$ to $\widehat{\omega}$ changes nothing at the right hand side that is because $\omega$ is co-closed and $\widehat{\omega}$ is closed
$$\nabla_{X}\rho=\frac{1}{p+1}i_{X}d(\omega+\widehat{\omega})-\frac{1}{n-p+1}\widetilde{X}\wedge d^{\dag}(\widehat{\omega}+\omega).$$
and the result (8) holds. This construction is represented in Figure 1, where $P$ stands for \textit{parallel forms}.
\def\setA{(0.8,0.8) ellipse (1.2cm and 1.2cm)}
 \colorlet{ellipse edge}{blue!50}
 \colorlet{ellipse area}{blue!20}

\def\setB{(-0.8,0.8) ellipse (1.2cm and 1.2cm)}
  \colorlet{ellipse edge}{red!50}
  \colorlet{ellipse area}{red!20}

\def\setC{(0.0,0.8) ellipse (2.3cm and 2.3cm)}
  \colorlet{ellipse edge}{green!50}
  \colorlet{ellipse area}{white!20}

\tikzset{filled/.style={fill=ellipse area, draw=ellipse edge, thick}, outline/.style={draw=ellipse edge, thick}}
\begin{center}
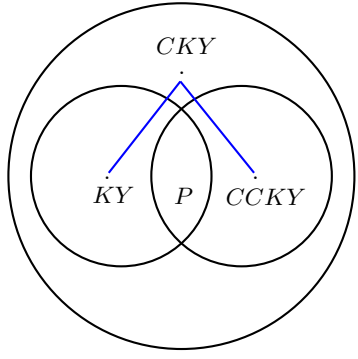
\begin{figure}
\begin{tikzpicture}[thick]

\begin{scope}

    \clip \setB;
    \fill[white!50] \setC;
      \end{scope}
      \begin{scope}

    \clip \setA;
    \fill[white!60] \setB;
\end{scope}

\draw \setA ;

\draw \setB
      node [xshift=-2.9,yshift=-7.0] {$KY$};
     \node[xshift=-28.00,yshift=22] {$\mathbf{.}$};
      {\draw [-][blue] (-0.966,0.844) -- (-0.014,2.06); }

\draw \setC ;

      {\node[xshift=0.9,yshift=15.4] {$P$} ; }

      \node[xshift=2,yshift=72] {$CKY$};
     \node[xshift=0.20,yshift=62] {$\mathbf{.}$};

      {\node[xshift=32,yshift=15.0] {$CCKY$} ; }
      \node[xshift=28.00,yshift=22] {$\mathbf{.}$};
      {\draw [-][blue] (-0.014,2.06) -- (0.966,0.844); }

\end{tikzpicture}
\caption{Building up a CKY form.}
\end{figure}
\end{center}
It is also possible to construct inhomogeneous self dual or anti-self dual CKY forms from a KY form or equivalently from a CCKY form. The sign is determined only by the degree of the unit Yano form \footnote{We use the term Yano form to indicate a KY form or a CCKY form, then the term \textit{dual Yano form} indicates a CCKY form or a KY form} because of the action of Hodge square on homogeneous forms; that is $$**\alpha=(-1)^{p(n-p)}\frac{det\mathbf{g}}{|det\mathbf{g}|}\alpha\,;\qquad \alpha \in \Gamma \Lambda^{p}(M),$$ here $\mathbf{g}$ is the chart matrix of the metric tensor $g$. If $\sigma\in\{\omega,\widehat{\omega}\}$ is a unit Yano form and if $**\alpha=\alpha$ then $\alpha+*\alpha$ is a self dual inhomogeneous CKY form; but if $**\alpha=-\alpha$ then $i\alpha+*\alpha$ is an anti-self dual inhomogeneous CKY form.\\\\ Here a \textit{physical interpretation} is inevitable according to the results of \cite{Acik Ertem 2015}; there the generalised Dirac currents (real homogeneous bilinears) of Killing spinors were satisfying a kind of higher degree Maxwell equations, so superposing the field associated to the co-existent Killing reversed spinors with the former field, a null higher-degree electromagnetic field for homogeneous CKY forms could be obtained. Passing to the inhomogeneous domain by linearity will turn out this null higher-degree electromagnetic field into a self dual or an anti-self dual one. The definition of $i=\sqrt{-1}$ is ambiguous here but since it exists at least if the Hodge map defines a complex structure on $\Gamma \Lambda^{p}(M)$ which is always possible for some $p$ \cite{Trautmann}.
 \subsection{Symmetry considerations}
 As mentioned before KY forms were used for the generation of first order symmetries of the massive Dirac equation $\displaystyle{\not}D\psi=m \psi$ as
\begin{eqnarray}
L_{(\omega)}:=\omega^{a}\nabla_{X_a}+\frac{p}{2(p+1)}d\omega
\end{eqnarray}
  namely $\displaystyle{\not}D L_{(\omega)}\psi= L_{(\omega)}\displaystyle{\not}D\psi=m L_{(\omega)}\psi$; $\omega^{a}:=i_{X^a}\omega$ and Clifford product is shown by the juxtaposition of the factors, but we put a dot when left Clifford acting on spinor fields. CKY forms corresponds similarly to massless Dirac equation $\displaystyle{\not}D\psi=0$ as
\begin{eqnarray}
L_{(\rho)}:=\rho^{a}\nabla_{X_a}+\frac{p}{2(p+1)}d\rho-\frac{n-p}{2(n-p+1)}d^{\dag}\rho
\end{eqnarray}
  i.e. $\displaystyle{\not}D L_{(\rho)}\psi=0$. Since $L_{(\rho)}=L_{(\omega+\widehat{\omega})}=L_{(\omega)}+\widehat{L}_{(\widehat{\omega})}$ then we can deduce that the first order operator generated by a CCKY form $\widehat{\omega}$ is
\begin{eqnarray}
\widehat{L}_{(\widehat{\omega})}=\widehat{\omega}^{a}\nabla_{X_a}-\frac{n-p}{2(n-p+1)}d^{\dag}\widehat{\omega}.
\end{eqnarray}
  \textit{The question here is, which equation admits this as its first order symmetry operator?} We are not going to answer this question here; but we should note that this operator does not reduce to the usual Lie derivative on spinor fields for degree one case, as opposed to $L_{(\omega)}$. The definition of $L_{(\rho)}$ for massless Dirac equation necessitates the existence of an operator $R$ \footnote{Confusion with the Riemann curvature $R$ tensor should be avoided.} such that
\begin{eqnarray}
\displaystyle{\not}D L_{(\rho)}=R\displaystyle{\not}D
\end{eqnarray}
i.e. $L_{(\rho)}$, $R$-\textit{commutes} with $\displaystyle{\not}D$. Although $\rho$, $\omega$ and $\widehat{\omega}$ are all homogeneous and of the same degree for our purposes, for general considerations they are taken as inhomogeneous forms which can be separated to their \textit{Clifford even} and \textit{Clifford odd} parts if needed. So, if $\Phi$ is one of the above Yano forms that is inhomogeneous, its even and odd parts respectively are $\Phi_{odd}=\sum_{p\,odd}\Phi_{p}$ and $\Phi_{even}=\sum_{p\,even}\Phi_{p}$ where $\Phi_{p}$ is the degree $p$ part of $\Phi$ obtained by applying the $p$-form projector $\wp_{p}$ on $\Phi$ \footnote{The properties associated to these projectors are completely given at the Appendix A of \cite{Acik Ertem 2015}}. For these general inhomogeneous investigations, the condition (14) transforms into
\begin{eqnarray}
\big[\displaystyle{\not}D, L_{(\rho)}\big]_{GCC}=R\displaystyle{\not}D
\end{eqnarray}
$[.\,,.]_{GCC}$ is the \textit{Graded Clifford Commutator} and the symmetry condition may be termed \textit{graded R-commuting}. Note that the odd and even parts of the first order symmetry operators satisfy  $$L'^{odd}_{(\Phi)}=L'_{(\Phi_{even})}\quad,\quad L'^{even}_{(\Phi)}=L'_{(\Phi_{odd})}\quad;\:L'\in\{L,\widehat{L}\}.$$ \textbf{A known result:} In any dimensions with arbitrary signature and curvature the below results hold \cite{Benn Kress2, Acik Ertem Onder Vercin1, Cariglia et. al.}.
\begin{itemize}
  \item The first order symmetry operator of an odd KY form Clifford commutes with the Dirac operator: $L_{(\omega_{odd})}\displaystyle{\not}D=\displaystyle{\not}DL_{(\omega_{odd})}.$
  \item The first order symmetry operator of an even CCKY form Clifford commutes with the Dirac operator: $\widehat{L}_{(\widehat{\omega}_{even})}\displaystyle{\not}D=\displaystyle{\not}D \widehat{L}_{(\widehat{\omega}_{even})}.$
  \item The first order operator of an even KY form Clifford anti-commutes with the Dirac operator: $L_{(\omega_{even})}\displaystyle{\not}D=-\displaystyle{\not}D L_{(\omega_{even})}.$
  \item The first order operator of an odd CCKY form Clifford anti-commutes with the Dirac operator: $\widehat{L}_{(\widehat{\omega}_{odd})}\displaystyle{\not}D=-\displaystyle{\not}D\widehat{L}_{(\widehat{\omega}_{odd})}.$
\end{itemize}
This result will be important for our homogeneous analysis, especially when the analog relations in the spinor sector will be derived. Here in the anti-commuting cases the first order operators are no more symmetry operators, but they in some sense have a physical meaning. In this case, if $\psi$ is a solution of the massive Dirac equation then $L\psi$ will be a solution of the Dirac equation with negative mass representing a hypothetical spinning particle. We will comment on this issue at Section III, but as every one knows that the quantum vacuum is full of these particles as part of pairs with energies of different signs permitted by the Heisenberg's quantum uncertainity principle.
\subsection{Physical interpretation and the main question}
In a recent work \cite{Acik Ertem 2015} we worked on the properties of bilinears generated by twistors and Killing spinors. The Killing spinor case, accompanied by a data set, was more sophisticated and rich. All possible outcomes obtainable from the Killing spinor bilinears were determined by the restrictive reality conditions imposed on them for physical reasons. As a reward we uncovered both kinematical and dynamical equations satisfied by the corresponding generalized Dirac currents of Killing spinor. The \textit{primitive set} of generating equations were seen to be
\begin{equation}
\nabla_{X_a} \sbl p= 2\lambda\, e_a \wedge \sbl {p-1}
\end{equation}
\begin{equation}
\nabla_{X_a} \sbl {p_{*}}= 2\lambda\, i_{X_a} \sbl {p_{*}+1},
\end{equation}
giving rise to the \textit{principal set}
\begin{equation}
d \sbl {p}=0 \qquad,\qquad d^{\dag}\sbl {p}=-2\lambda (n-p+1) \sbl {p-1}\:;
\end{equation}
\begin{equation}
d\sbl {p_{*}}= 2\lambda (p_{*}+1) \sbl {p_{*}+1}\qquad,\qquad d^{\dag}\sbl {p_{*}}=0.
\end{equation}
Here $p_{*}$ means that it has a different parity than $p$, i.e. $p_{*}+p$ is always odd. These equations imply that the homogeneous realified parts of Killing spinor bilinears represent Duffin-Kemmer-Petiau (DKP) fields or Maxwell-like fields on the dynamical side, where on the kinematical side when the degree $p$ part of the bilinear satisfies KY equation the degree $p_{*}$ part satisfies CCKY equation and vice-versa. The kinematical results could then be applied by using the machinery of the previous subsection. Namely for a Killing spinor $\psi$, we know that for some $p$, $\sbl {p}$ is a KY form which requires that $\sbl {p_{*}}$ is a CCKY form then we also know that $\sbl {p}+\sbl {p_{*}}$ is a CKY form which is generally associated to a twistor spinor's bilinear. Now the remarkable question arises!

\begin{flushleft}
\textbf{The main question:} \textit{By using the generalized currents of Killing spinors one can deduce directly the generalized currents of twistors. Is there a way for generating twistors from Killing spinors? } \\
\end{flushleft}

Before passing to the next section in search for the answer, we want to repeat a previous interpretation of the equations (14) and (15). With a slight change of understanding the primitive equations could be reinterpreted as follows: The propagation of a brane in spacetime is triggered by the creation of a brane with one lower dimension and because of the unstable motion of the higher dimensional brane it is annihilated and this gives rise to the propagation of the lower dimensional (stable) brane. Of course in the context of General Relativity this process should be observed in a locally inertial frame $\{X_a\}$ co-moving with the stable brane to which it is adapted in a such way that $g(X_0,X_0)+\sum_{i}g(X_i,X_i)<0$ and this corresponds to the physical (semi-classical) motion of this system (for details see the insight in \cite{Acik Ertem 2015}). Some classical references for the details of the motion of extendons in General Relativity are \cite{Geroch Traschen, Stachel a, Stachel b, Tucker mem, Tucker et al}. The selection of the local frame is important for the preservation of local causality which temporally orders the equations in the primitive set in accordance with the above scenario; otherwise the mathematical simultaneity of the equations will be deceptive. The temporal parameter should be taken as the local proper time measured by the locally inertial (time-like) observer instead of the local coordinate time. If one remembers that $e^a\wedge$ is equal to ${i_{X_a}}^{\dag}:=*^{-1}i_{X_a}*\eta$ (see App. A of \cite{Acik Ertem 2015}), the primitive set could be rewritten as
\begin{equation}
{i_{X_a}}^{\dag} \sbl {p-1} = (2\lambda)^{-1}\,\nabla_{X_a}\sbl p  \nonumber
\end{equation}
\begin{equation}
i_{X_a} \sbl {p_{*}+1}= (2\lambda)^{-1}\,\nabla_{X_a} \sbl {p_{*}} , \nonumber
\end{equation}
those contain more physical intuition. Our scenario could also be thought as equivalent to Dirac's \cite{Dirac a} where he models the electron as an extended elementary particle and the muon corresponds to the first excited state of the electron; a similar and more accurate construction may be found in \cite{Onder Tucker}. From this point of view, our model may be seen as a gas of one level $(p-2)$-branes (i.e. the $(p-2,p-1)$-brane couple), if $n$ is the dimension of spacetime then $p$ ranges from $2,4,...,n+2$ in odd dimensions and from $2,4,...,n$ in even dimensions. This set up requires one level branes because $i_{X_a}$ and ${i_{X_a}}^{\dag}$ are both nilpotent of index two.

\section{Building up twistors from Killing spinors}
\subsection{The fermionic sector}
The answer to the main question is positive. If $\psi$ is a Killing spinor
\begin{equation}
\nabla_{X}\psi=\lambda \widetilde{X}.\psi
 \end{equation}
then we define the Killing reversal $\psi^{\varsigma}$ of $\psi$ by
\begin{equation}
\nabla_{X}\psi^{\varsigma}=-\lambda \widetilde{X}.\psi^{\varsigma}\;,
\end{equation}
that was defined in \cite{Acik JMP 2016} technically, but is being known and used for example in \cite{Fujii Yamagishi}. The generation of twistors from Killing spinors is given by the following proposition.\\

\textbf{Proposition:} To every Killing spinor pair $\psi,\psi^{\varsigma}$ there corresponds a twistor pair $\Psi^{+},\Psi^{-}$ such that $$\Psi^{\pm}=\psi\pm\psi^{\varsigma},$$
where $\psi^{\varsigma}$ is the Killing reversal of $\psi$; and trivially the Killing reversals of the induced twistors are ${\Psi^{\pm}}^{\varsigma}=\pm\Psi^{\pm}$.\\

\textbf{Proof:} $$\nabla_{X_a}\Psi^{\pm}=\nabla_{X_a}(\psi\pm\psi^{\varsigma})=\nabla_{X_a}\psi\pm\nabla_{X_a}\psi^{\varsigma}=\lambda e_{a}.\psi\mp\lambda e_{a}.\psi^{\varsigma}=\lambda e_{a}.\Psi^{\mp},$$ and if we Clifford contract both hand sides from left with $e^{a}$ and use the identity $e^a e_a=n$ then we obtain $$\displaystyle{\not}D\Psi^{\pm}=n\lambda \Psi^{\mp}$$ and if we put the last identity into the last equality of the first relation we reach the desired result $$\nabla_{X_a}\Psi^{\pm}=\frac{1}{n}e_{a}.\displaystyle{\not}D\Psi^{\pm},$$ namely the twistor equation.\\

\begin{flushleft}
\textbf{Physical Interpretation:}
\end{flushleft}
The process $\psi+\psi^{\varsigma}\rightarrow\Psi^{+}+\Psi^{-}$ mimics the well known quantum electrodynamical pair annihilation process $e^{-}+e^{+}\rightarrow \gamma+\gamma$ in many ways. From sections II.C and II.D we know that Killing spinors are related to the massive sector and the twistors are related to the massless sector; so just as the mutual annihilation of an electron and a positron results in a pure electromagnetic field the mutual annihilation of a Killing spinor and its Killing reversal results in a pure radiation field the kind of which should be determined by further considerations. This can partially be achieved by the comparison of the properties of the Killing reversal map with that of charge conjugation (mainly) for the complex case. Another possibly related interpretation could be that the $\{\psi,\psi^{\varsigma}\}$ pair is extracted from the quantum vacuum by the intense gravitational field of a black hole around the horizon which conceptually resembles the gravitationally sourced thermal radiation of a collapsed body into a black hole, namely the Hawking radiation. So in our model, the outward radiation of a Killing spinor field shall be termed the \textit{Hawking-Killing radiation} of a black hole. Whereas the evaporation is due to the inward motion of the negative energy reversed Killing spinor field. Also, the tie with Wigner's time reversal may also be checked locally because of the relation between charge conjugation and time reversal; this at a first investigation requires the usage of locally Minkowskian nature of spacetime. Since the resultant photons have different helicities, the produced twistors also should have different helicity states for the consistency of our analog model.\\
\begin{flushleft}
 \textit{\underline{Charge conjugation versus Killing reversal}}:
\end{flushleft}
The similarities and differences between charge conjugation and Killing reversal is given in Table 1. Since the charge conjugation in some sense carries the properties of electromagnetic interactions then by comparison the interaction to which Killing reversal belongs to could be worked out.
\begin{table}[h]
\centering
\begin{tabular}{c|c}

  $ Charge\,Conjugation$ & $Killing\,Reversal$  \\ \hline
  $(\psi^{c})^{c}=\pm \psi$ & $(\psi^{\varsigma})^{\varsigma}=\psi$ \\
  $(\nabla_{X}\psi)^{c}=\nabla_{X}{\psi^{c}}$ & $(\nabla_{X}\psi)^{\varsigma}\neq\nabla_{X}{\psi^{\varsigma}}$ \\
  $(\pounds_{K}\psi)^{c}\neq \pounds_{K}{\psi^{c}}$ & $(\pounds_{K}\psi)^{\varsigma} = \pounds_{K}{\psi^{\varsigma}}$ \\
\end{tabular}
\caption{Properties of charge conjugation and Killing reversal maps.}
\end{table}
 The behaviour of charge conjugation depends on the dimension and signature but Killing reversal is independent from them; also while acting on real spinors charge conjugation maps a spinor $\psi$ to $\pm\psi$ on the other hand Killing reversal maps both real and complex spinors in the same manner. Charge conjugation and Killing reversal both preserve the spin degrees of freedom, charge conjugation changes the sign of the electric charge whereas the Killing reversal changes the sign of the inertial mass or more generally energy. This could be seen if one Clifford contracts the Killing spinor equation from the left $e^{a}.\nabla_{X_a}\psi=\lambda e^{a}e_{a}.\psi$ i.e.
\begin{equation}
\displaystyle{\not}D\psi=\lambda n\psi\,,
\end{equation}
and as a result identifies $n.\lambda$ with the inertial mass $m$. From this point of view the Killing reversal map may be associated to pure gravitational interactions. Here $\pounds_{K}$ is the spinorial Lie derivative with respect to a Killing vector field and it sends a Killing spinor to a Killing spinor so it commutes with the Killing reversal map. Similarly the  first order operator with respect to a odd KY form $L_{(\omega)}$ is a symmetry operator for Killing spinor equation hence should commute with the Killing reversal i.e. $(L_{(\omega)}\psi)^{\varsigma}=L_{(\omega)}\psi^{\varsigma}$. Exact coincidences between these operations could be possible when a gravitational problem is reducible to an electromagnetic one in curved spacetime quantum field theories \cite{Fulling}.\\
\begin{flushleft}
\textit{\underline{Comment on the relation to Wigner time reversal}}:
\end{flushleft}
Discrete orientation-changing finite diffeomorphisms could be worked out locally in a general curved spacetime, so then the only guide will be the local validity of special relativity at a first approximation. But if one intends to analyze the higher order effects of the diffeomorphism flow in a curved spacetime generated by the Wigner time reversal operation, which is closely related to charge conjugation in a flat spacetime, the past-future asymmetry of the gravitational field would avoid a well-defined analysis. Furthermore, the requirement of global hyperbolicity on spacetime could pose a well-behaved ''time'' analysis and then its relation to other operations such as charge conjugation and Killing reversal shall be determined. In a flat spacetime covariances of Dirac equation (possibly coupled to a Maxwell field) under the above mentioned finite isometric diffeomorphisms are formed by labelling the later by a parallel element $s$ of the smooth sections of Clifford group bundle $\bigcup_{m\in M}\Gamma_{m}(M,\eta)$ where $\Gamma_{m}(M,\eta)$ is the Cifford group at the point $m\in M$. In the presence of a preferred inner product for spinor fields, the Clifford group bundle should be restricted to the invariance group bundle of the spinor product \cite{Benn Tucker}.\\
\begin{flushleft}
\textit{\underline{Helicity considerations}}:
\end{flushleft}
The analogy made before makes one to anticipate that the resultant twistor fields should correspond to different helicity states. If we define $\Psi:= \Psi^{+}+\Psi^{-}$ and if we restrict ourselves to even dimensional Lorentzian spacetimes for physical reasons then we can represent the Killing reversal operation by the left action of the volume form $z:=*1$ on a Killing spinor or on a twistor induced from a Killing spinor. If $z^2=1$ the helicity operators are $\frac{1}{2}(1\pm z)$ and if $z^2=-1$ then the operator in this case are $\frac{1}{2}(1\pm iz)$; note that in even dimensions Clifford algebras are central simple and $z$ Clifford anti-commutes with odd forms. When $z^2=-1$ the center is algebraically isomorphic to $\mathbb{C}$ and hence the helicity operators are well defined. Lets see that $\psi^{\varsigma}=z.\psi$ under our assumptions: If $\nabla_{X}\psi=\lambda \widetilde{X}.\psi$ then $\nabla_{X}z.\psi=z.\nabla_{X}\psi=\lambda z\widetilde{X}.\psi=-\lambda \widetilde{X}z.\psi$, so it is legitimate to identify both. Let us take $z^2=1$ then $$\frac{1}{2}(1\pm z)\Psi=(1\pm z)\psi=(\psi\pm z.\psi)=(\psi\pm \psi^{\varsigma})=\Psi^{\pm},$$ which ends the proof. It is interesting to note here that when the main anti-automorphism $\xi$ of the Clifford algebra corresponds to the adjoint involution of the spinor inner product then the \textit{Hodge dual of a spinor} is well defined and it maps a spinor that is an element of a Minimal Left Ideal to a dual spinor i.e. an element of the associated Minimal Right Ideal. This follows from the well known identity for Clifford forms namely the \textit{Clifford-K\"{a}hler-Hodge duality} $*\Phi=\Phi^{\xi} z$; so $$\nabla_{X}*\psi=*\nabla_{X}\psi=\lambda *(\widetilde{X}.\psi)=\lambda (\widetilde{X}.\psi)^{\xi}.z=\lambda \overline{\psi}.\widetilde{X}^{\xi}z=\lambda \overline{\psi}.*\widetilde{X}$$ that means $*\psi$ is the associated dual spinor sharing the same Killing number with $\psi$.\\\\
\textbf{Geometric Identities}: The action of the Hessian $\nabla^{2}(X_a,X_b):=\nabla_{X_a}\nabla_{X_b}-\nabla_{\nabla_{X_a}X_b}$ on a Killing spinor $\psi$ is $$\nabla^{2}(X_a,X_b)\psi=\lambda^{2}e_{ba}.\psi$$ and $$\mathbf{R}(X_a,X_b)\psi=\lambda^{2}(e_{ba}-e_{ab}).\psi=2\lambda^{2}(e_{b}\wedge e_{a}).\psi$$ or it can be rewritten as $\mathbf{R}(X_a,X_b)\psi=\lambda^{2}e_{ba}.\psi; \:a\neq b$. The trace of the Hessian is defined by $\nabla^2:=\nabla^{2}(X_a,X^a)=n\lambda^{2}\psi$, so since by \textit{Schr\"{o}dinger-Weitzenb\"{o}ck-Bochner-Lichnerowicz formula} for spinors, the square of the Dirac operator is
\begin{eqnarray}
  \displaystyle{\not}D^{2}\psi &=& \nabla^2\psi-\frac{1}{2}\psi.\mathbf{R}(X_a,X_b)e^{ba} \\ \nonumber
  &=& \nabla^2\psi-\frac{1}{4}\psi. R_{ab} e^{ba} \nonumber
  = \nabla^2\psi-\frac{1}{4}\mathcal{R}\psi.  \nonumber
\end{eqnarray}
Finally from the geometric constraint $\mathcal{R}=-4\lambda^{2}n(n-1)$ for the existence of Killing spinors we have
\begin{eqnarray}
  \displaystyle{\not}D^{2}\psi = \lambda^{2}n^{2}\psi.
\end{eqnarray}
A trivial result which could be directly obtained from (20).

\subsection{The bosonic sector}
The inhomogeneous Clifford forms constructed from the induced twistors decompose into the bilinears of generator Killing spinors as $$\Psi^{\pm}\overline{\Psi^{\pm}}=\psi\overline{\psi}+\psi^{\varsigma}\overline{\psi^{\varsigma}}\pm\psi^{\varsigma}\overline{\psi}\pm\psi\overline{\psi^{\varsigma}},$$
the primitive set of equations corresponding to each of them, and their principal sets together with their determining Clifford algebraic operators  are\\
I
\begin{eqnarray}
 \nabla_{X_a} \sbl p &=& 2\lambda\, e_a \wedge \sbl {p-1},\qquad \widehat{\lambda}^{\pm}_{p} \\ \nonumber
\nabla_{X_a} \sbl {p_{*}} &=& 2\lambda\, i_{X_a} \sbl {p_{*}+1}; \nonumber
\end{eqnarray}
\begin{eqnarray}
d \sbl {p}&=&0 \qquad,\qquad d^{\dag}\sbl {p}=-2\lambda (n-p+1) \sbl {p-1}\:; \\ \nonumber
d\sbl {p_{*}}&=& 2\lambda (p_{*}+1) \sbl {p_{*}+1}\qquad,\qquad d^{\dag}\sbl {p_{*}}=0,\nonumber
\end{eqnarray}
\underline{------------------------------------------------------------------------------------------------------------------------------------------------------------}
II
\begin{eqnarray}
\nabla_{X_a} (\psi^{\varsigma}\overline{\psi^{\varsigma}})_{p}&=& -2\lambda\, e_a \wedge (\psi^{\varsigma}\overline{\psi^{\varsigma}})_{p-1},\qquad -\widehat{\lambda}^{\pm}_{p}\\ \nonumber
\nabla_{X_a} (\psi^{\varsigma}\overline{\psi^{\varsigma}})_{p_{*}}&=& -2\lambda\, i_{X_a} (\psi^{\varsigma}\overline{\psi^{\varsigma}})_{p_{*}+1};\nonumber
\end{eqnarray}
\begin{eqnarray}
d \sbl {p}&=&0 \qquad,\qquad d^{\dag}\sbl {p}=2\lambda (n-p+1) \sbl {p-1}\:; \\ \nonumber
d\sbl {p_{*}}&=& -2\lambda (p_{*}+1) \sbl {p_{*}+1}\qquad,\qquad d^{\dag}\sbl {p_{*}}=0,\nonumber
\end{eqnarray}
\underline{------------------------------------------------------------------------------------------------------------------------------------------------------------}
III
\begin{eqnarray}
\nabla_{X_a} (\psi^{\varsigma}\overline{\psi})_{p}&=& 2\lambda\, i_{X_a} (\psi^{\varsigma}\overline{\psi})_{p+1},\qquad \widehat{\lambda}^{\mp}_{p}\\ \nonumber
\nabla_{X_a} (\psi^{\varsigma}\overline{\psi})_{p_{*}}&=& 2\lambda\, e_a \wedge (\psi^{\varsigma}\overline{\psi})_{p_{*}-1};\nonumber
\end{eqnarray}
\begin{eqnarray}
d\sbl {p}&=& 2\lambda (p+1) \sbl {p+1}\qquad,\qquad d^{\dag}\sbl {p}=0; \\ \nonumber
d \sbl {p_{*}}&=&0 \qquad,\qquad d^{\dag}\sbl {p_{*}}=-2\lambda (n-p_{*}+1) \sbl {p_{*}-1}\:,\nonumber
\end{eqnarray}
\underline{------------------------------------------------------------------------------------------------------------------------------------------------------------}
IV
\begin{eqnarray}
\nabla_{X_a} (\psi\overline{\psi^{\varsigma}})_{p}&=& -2\lambda\, i_{X_a} (\psi\overline{\psi^{\varsigma}})_{p+1}, \qquad -\widehat{\lambda}^{\mp}_{p} \\ \nonumber
\nabla_{X_a} (\psi\overline{\psi^{\varsigma}})_{p_{*}}&=& -2\lambda\, e_a \wedge (\psi\overline{\psi^{\varsigma}})_{p_{*}-1};\nonumber
\end{eqnarray}
\begin{eqnarray}
d\sbl {p}&=& -2\lambda (p+1) \sbl {p+1}\qquad,\qquad d^{\dag}\sbl {p}=0; \\ \nonumber
d \sbl {p_{*}}&=&0 \qquad,\qquad d^{\dag}\sbl {p_{*}}=2\lambda (n-p_{*}+1) \sbl {p_{*}-1}\:,\nonumber
\end{eqnarray}
here $\hat{\lambda}^{\pm}_{(p)}=(\lambda 1 \pm (-1)^{p} \lambda^{j_c} \mathcal{J})$ with $\mathcal{J}$ the adjoint involution of the spinor inner product and $j_c$ the induced involution on the real algebra of complex numbers. The first set pairs I and II do have the same Yano type and the last set pairs III and IV either, but first and last have different types. In fact the twistor bilinears satisfy CKY equation \cite{Acik Ertem 2015}, but an alternative proof follows from calculating their covariant derivatives and using the Yano properties of the Killing spinor bilinears. Another thing is that, if one tries to obtain the primitive or principal sets associated to twistor bilinears from those of Killing spinor bilinears, then he/she should remember that while the former ones are free from the reality conditions the latter ones are not.\\\\
\textbf{Geometric Identities}: \\
\textbf{(a)} From (14) the Hessian of $(\psi\overline{\psi})_{p}$ is found to be
\begin{eqnarray}
\nabla^{2}(X_a,X_b)(\psi\overline{\psi})_{p}=4\lambda^{2}e_{b}\wedge i_{X_a}(\psi\overline{\psi})_{p}=4\lambda^{2}i_{X_{b}}^{\dag}i_{X_a}(\psi\overline{\psi})_{p}
\end{eqnarray}
and necessarily
\begin{eqnarray}
\mathbf{R}(X_a,X_b)(\psi\overline{\psi})_{p}=-4\lambda^{2}(i_{X_{a}}^{\dag}i_{X_b}-i_{X_{b}}^{\dag}i_{X_a})(\psi\overline{\psi})_{p}.
\end{eqnarray}
This together with $$\nabla^{2}(\psi\overline{\psi})_{p}=4\lambda^{2}p(\psi\overline{\psi})_{p}$$ gives
\begin{eqnarray}
\displaystyle{\not}d^{2}\sbl {p}&=& \nabla^{2}(\psi\overline{\psi})_{p}-\frac{1}{2}\mathbf{R}(X_a,X_b)\sbl {p}e^{ab} \\ \nonumber
&=&4\lambda^{2}p(\psi\overline{\psi})_{p}+2\lambda^{2}\big((i_{X_{a}}^{\dag}i_{X_b}-i_{X_{b}}^{\dag}i_{X_a})(\psi\overline{\psi})_{p}\big) e^{ab}\\ \nonumber
&=&4\lambda^{2}p(n-p+1)(\psi\overline{\psi})_{p}.\nonumber
\end{eqnarray}
The derivation of the equations (31) and (32) can be found in Appendix C.\\\\
\textbf{(b)} Similar identities for (15) are as follows:
\begin{eqnarray}
\nabla^{2}(X_a,X_b)(\psi\overline{\psi})_{p_*}=4\lambda^{2}(\eta_{ab}-e_{a}\wedge i_{X_b})(\psi\overline{\psi})_{p_*},
\end{eqnarray}
so
\begin{eqnarray}
\mathbf{R}(X_a,X_b)(\psi\overline{\psi})_{p_*}=-4\lambda^{2}(i_{X_{a}}^{\dag}i_{X_b}-i_{X_{b}}^{\dag}i_{X_a})(\psi\overline{\psi})_{p_*}.
\end{eqnarray}
Unifying the last one with $$\nabla^{2}(\psi\overline{\psi})_{p_*}=4\lambda^{2}(n-p_*)(\psi\overline{\psi})_{p_*}$$ leaves us with
\begin{eqnarray}
\displaystyle{\not}d^{2}\sbl {p_*}&=&4\lambda^{2}(p_*+1)(n-p_*)(\psi\overline{\psi})_{p_*}.
\end{eqnarray}
As a result (31) and (32) are homogeneous equations in the form of eigenvalue equations of the Laplace-Beltrami operator, and are coupled in accordance with (18) and (19). Note that when $p_{*}=p-1$ then the coupled fields do have the same mass.

 \section{Conclusion}

The aim of our programme is to work out the many details for putting Killing spinors into the main elements of mathematical physics. The start was given in \cite{Acik Ertem 2015}, became manifest in \cite{Acik JMP 2016} where also a road map was constructed. This latter work contained some open questions, each of which could well be a problem of its own. The present work fills one of these important gaps theoretically and promises a companion paper for the application of its results; the plan of this second part will be explained below. Before that we want to make a brief report about the new contributions to the literature by this work.\\

We first gave a new elegant form for defining KY forms and CCKY forms, that is complementing the esthetic inhomogeneous definitive equation for CKY forms which was given before in \cite{Acik Ertem 2015}; so a trivial but effective way for building up a CKY form out of a KY form and a CCKY form
was at hand. This simple consequence made us identify clearly the corresponding first order symmetry operators for massive and massless Dirac equations in curved spacetime admitting KY and CCKY forms, removing some ambiguities existent in the literature. By the way we showed that one can set up self-dual or anti self-dual massless fields from the resultant CKY forms, this result is in accordance with the analogies given in \cite{Acik Ertem 2015}. Some of the above mentioned operators anti-commuting with Dirac operator reveal the negative energy massive spinning particles, at first sight seen as unphysical or virtual but the value of this became apparent later on. Again from previous publications it was known that the generalised Dirac currents of Killing spinors were identified with KY and CCKY forms on the kinematical side; so we asked the possibility for generating twistors from Killing spinors and reached the positive answer by using some original technical details. The most critical physical interpretation was that the generation of a twistor pair by the Killing spinor pair, was identified with the quantum electrodynamical pair annihilation process and also eventually to the Hawking radiation of a black hole. The characteristic mathematical operation associated to this analogy which is termed the Killing reversal was compared and related to the charge conjugation and time-reversal for completeness. Also the investigations in the fermionic sector uncovered the relation of Killing number to inertial mass which is dimension dependent; this relates the dimensionality of space-time to the concept of inertia in some sense. The analysis of the bosonic part added three more types of primitive and principal sets of equations to the preceding ones. A thing not to pass without mentioning is the translation of the coordinate-free form of the primitive set of equations to the coordinate-wise ones, given in Appendix C. Another idea could be to use the reverse procedure for obtaining Killing spinors from twistors in spacetimes admitting the latter ones by the method of trial and error roughly. We also emphasized the fact that our technique shall be of use in classifying spacetimes admitting twistors or Killing spinors.\\

The planed companion work called Part II \cite{Acik 2017} will be dealing with many topics. These include the stress tensors of the specific spinor fields and the stress tensors of their generalised Dirac currents, the continuation and improvement of the quantum field theoretical formulation of physical examples given in \cite{Acik JMP 2016} and the present work, selection of a black hole spacetime admitting a Killing spinor for clarifying the Hawking-Killing radiation process, fixing the values of the inherent degrees associated to the dimensions of the brane immersions including strings and membranes and then reconsidering the dynamical equations such as DKP and Maxwell-like equations in this context. Last but not least we hope to calculate the Killing spinor existing in the plane-wave spacetime and push the button for using our method again.

\appendix
\begin{flushleft}
\section{Evaluation of ${L_{(\omega)}}^{\dag}$ and $\widehat{L}_{(\widehat{\omega})}^{\dag}$ and a comment on the symmetry algebra}
\end{flushleft}
Recall that from (11), $L_{(\omega)}:=\omega^{a}\nabla_{X_a}+\frac{p}{2(p+1)}d\omega$ so from metric compatibility of the connection one can write $${L_{(\omega)}}^{\dag}=(\omega^{a})^{\dag}\nabla_{X_a}+\frac{p}{2(p+1)}(d\omega)^{\dag}=*^{-1}i_{X_{a}}\omega{*}\eta\nabla_{X_a}+\frac{p}{2(p+1)}*^{-1}d\omega{*}\eta$$
and also using the identity $*^{-1}i_{X}\Phi=(*^{-1}\Phi)\wedge \widetilde{X}$, $${L_{(\omega)}}^{\dag}=\bigg(\big((*^{-1}\omega)\wedge e^{a}\big)\nabla_{X_a}+\frac{p}{2(p+1)}d^{\dag}\eta*^{-1}\omega\bigg){*}\eta=\big((e^{a}\wedge \eta*^{-1}\omega)\nabla_{X_a}+\frac{p}{2(p+1)}d^{\dag}\eta*^{-1}\omega\big){*}\eta.$$ Finally after using the identity $*^{-1}\omega=(-1)^{p(n-p)}\varepsilon(\mathbf{g})*\omega$, so $${L_{(\omega)}}^{\dag}=(-1)^{(p+1)(n-p)}\varepsilon(\mathbf{g})\big((e^{a}\wedge *\omega)\nabla_{X_a}+\frac{p}{2(p+1)}d^{\dag}*\omega\big){*}\eta.$$ Let us define $\widehat{\widehat{L}}_{(*\omega)}:=((e^{a}\wedge *\omega)\nabla_{X_a}+\frac{p}{2(p+1)}d^{\dag}*\omega\big)$ then $${L_{(\omega)}}^{\dag}=(-1)^{(p+1)(n-p)}\varepsilon(\mathbf{g})\widehat{\widehat{L}}_{(*\omega)}{*}\eta,$$ here $\varepsilon(\mathbf{g}):=\frac{det(\mathbf{g})}{|det(\mathbf{g})|}$ where $\mathbf{g}$ is the chart matrix of the metric tensor. The last relation could also be written as
\begin{eqnarray}
*{L_{(\omega)}}=\widehat{\widehat{L}}_{(*\omega)}=((e^{a}\wedge *\omega)\nabla_{X_a}+\frac{p}{2(p+1)}d^{\dag}*\omega\big).
\end{eqnarray}
Another thing to be done for completeness is to calculate $*\widehat{L}_{\widehat{\omega}}$, and with a little algebra it is found that
\begin{eqnarray}
*\widehat{L}_{\widehat{\omega}}=(-1)^{(n-p+1)}\big((e^a\wedge *\widehat{\omega})\nabla_{X_a}+\frac{n-p}{2(n-p+1)}d*\widehat{\omega}\big).
\end{eqnarray}
\\Dimension dependent closure of the symmetry algebra is based on the set of odd KY forms and even CCKY forms, then the associated first order operators $L^{even}_{KY}$'s and $\widehat{L}^{odd}_{CCKY}$ 's form an algebra under Killing-Yano brackets (ref. Cariglia et. al.). In the last reference a detailed account of symmetry analysis could be found, but there are some sign ambiguities arising from minor errors.\\
\section{Some general relations between $d$ and $\nabla$}
The word \textit{general} in the title means that we may be considering connections with torsion in our analysis in addition to our Riemannian considerations throughout the text. Let us specify three kinds of metric compatible connections for our purposes: the Levi-Civita connection $\nabla^{LC}$, non-Riemannian connection with torsion $\nabla^{NR}$ and a variable connection $\nabla \in \{\nabla^{LC}, \nabla^{NR}\}$. Although Cartan's exterior derivative $d$ is a product(ion) of the \textit{differentiable structure} of the spacetime manifold, the compatibility with the \textit{notion of a parallelism} makes it possible to be written respectively as $d=e^{a}\wedge \nabla^{LC}_{X_a}$ and $d=e^{a}\wedge \nabla^{NR}_{X_a}+T^{a}\wedge i_{X_a}$. If $\Omega \in \Gamma\Lambda^{p}(M)$ and $Y_{i}\in \Gamma TM; i=1,2,...,p+1$ then by using (3) it is easy to show by induction that
\begin{eqnarray}
(e^{a}\wedge \nabla_{X_a} \Omega)(Y_{1}, Y_{2}, ..., Y_{p+1})=\frac{1}{(p+1)!}\sum_{k=0}^{p}(-1^{kp}) \Pi_{k}(i_{Y_{p+1}}...i_{Y_{2}}\nabla_{Y_{1}}\Omega)
\end{eqnarray}
where $\Pi_{k}=(\Pi_{1})^{k}$ with $$\Pi_{1}(i_{Y_{p+1}}...i_{Y_{2}}\nabla_{Y_{1}}\Omega)=i_{Y_{p}}...i_{Y_{2}}i_{Y_{1}}\nabla_{Y_{p+1}}\Omega.$$ $\Pi_{0}$ is the identity permutation and $\Pi_{1}$ is the cyclic permutation $\Pi_{1}=\binom{\,i_{1}i_{2}...i_{p+1}}{i_{2} i_{3}\, ...\, i_{1}}$ which together with transpositions generate the whole permutation group $S(p+1)$. Again using (3) but in reverse order, (B1) can be rewritten as
\begin{eqnarray}
(e^{a}\wedge \nabla_{X_a} \Omega)(Y_{1}, Y_{2}, ..., Y_{p+1})=\frac{1}{(p+1)}\sum_{k=0}^{p}(-1^{kp}) \Pi_{k}((\nabla_{Y_{1}}\Omega)(Y_{2},..., Y_{p+1} ))
\end{eqnarray}
 or from the definition of the covariant differential this becomes
\begin{eqnarray}
(e^{a}\wedge \nabla_{X_a} \Omega)(Y_{1}, Y_{2}, ..., Y_{p+1})&=&\frac{1}{(p+1)}\sum_{k=0}^{p}(-1^{kp}) \Pi_{k}((\nabla\Omega)(Y_{1}, Y_{2},..., Y_{p+1} ))\\ \nonumber &=&\frac{1}{(p+1)}\sum_{k=0}^{p}(-1^{kp}) \Pi_{k}((\nabla\Omega)(e^{a}(Y_{1})X_a, Y_{2},..., Y_{p+1} )) \\ \nonumber
&=&\frac{1}{(p+1)}\nabla\Omega\bigg(\sum_{k=0}^{p}(-1^{kp}) \Pi_{k}((e^{a}(Y_{1})X_a, Y_{2},..., Y_{p+1} ))\bigg).  \nonumber
\end{eqnarray}
\textbf{Example ($p=2$)}: For $F \in \Gamma \Lambda^{2}(M)$ and $X,Y,Z\in \Gamma TM$, let us evaluate the above identity specifically.
\begin{equation}
(e^{a}\wedge \nabla_{X_a} F)(X,Y,Z)=\frac{1}{3}\big((\nabla F)(X,Y,Z)-(\nabla F)(Y,Z,X)+(\nabla F)(Z,X,Y)\big).\\ \nonumber
\end{equation}
\section{The coordinate expressions of the primitive set of equations}
This appendix is intended to make clear the understanding of the basic equations of our programme, for the general reader who is familiar with the more common notation based on local components of tensor fields.\\
We only translate the primitive set of equations. Let us define $\sbl {p}:=\Omega_{p}$ for brevity and work in a local chart with coordinate functions $x=(x^{\mu})$. Since our equations are general covariant we can write, for example (16) as:
\begin{equation}
\nabla_{\frac{\partial}{\partial x^{\mu}}}\Omega_{p}=2\lambda dx_{\mu}\wedge \Omega_{p-1}. \nonumber
\end{equation}
The local expansions of the form fields are $$\Omega_{p}=\frac{1}{p!}(\Omega_{p})_{\sigma_{1}...\sigma_{p}}dx^{\sigma_{1}}\wedge...\wedge dx^{\sigma_{p}}$$ and $$\Omega_{p-1}=\frac{1}{(p-1)!}(\Omega_{p})_{\sigma_{1}...\sigma_{p-1}}dx^{\sigma_{1}}\wedge...\wedge dx^{\sigma_{p-1}}.$$
So,
\begin{equation}
\nabla_{\frac{\partial}{\partial x^{\mu}}}\big((\Omega_{p})_{\sigma_{1}...\sigma_{p}}dx^{\sigma_{1}}\wedge...\wedge dx^{\sigma_{p}}\big)=2\lambda p (\Omega_{p})_{\sigma_{1}...\sigma_{p-1}}dx_{\mu}\wedge dx^{\sigma_{1}}\wedge...\wedge dx^{\sigma_{p-1}}, \nonumber
\end{equation}
the left hand side is
\begin{eqnarray}
 \nonumber % Remove numbering (before each equation)
&\nabla&_{\frac{\partial}{\partial x^{\mu}}}\big((\Omega_{p})_{\sigma_{1}...\sigma_{p}}dx^{\sigma_{1}}\wedge...\wedge dx^{\sigma_{p}}\big)=(\partial_ {\mu}(\Omega_{p})_{\sigma_{1}...\sigma_{p}})dx^{\sigma_{1}}\wedge...\wedge dx^{\sigma_{p}}+(\Omega_{p})_{\sigma_{1}...\sigma_{p}}\nabla_{\frac{\partial}{\partial x^{\mu}}}(dx^{\sigma_{1}}\wedge...\wedge dx^{\sigma_{p}}) \\ \nonumber
&=&(\Omega_{p})_{\sigma_{1}...\sigma_{p},\mu}dx^{\sigma_{1}}\wedge...\wedge dx^{\sigma_{p}}-{\omega^{\sigma_{i}}}_{\kappa}(\partial_{\mu})(\Omega_{p})_{\sigma_{1}...\sigma_{p}}(dx^{\sigma_{1}}\wedge...dx^{\sigma_{i-1}}\wedge dx^{\sigma_{\kappa}}\wedge dx^{\sigma_{i+1}}\wedge...\wedge dx^{\sigma_{p}}) \\ \nonumber
&=&\big((\Omega_{p})_{\sigma_{1}...\sigma_{p},\mu}-\sum_{i}{\Gamma_{\mu\,\sigma_{i}}}^{\kappa} (\Omega_{p})_{\sigma_{1}...\sigma_{i-1}\kappa \sigma_{i+1}...\sigma_{p}}\big)dx^{\sigma_{1}}\wedge...\wedge dx^{\sigma_{p}} \nonumber
\end{eqnarray}
conventionally written componentwise as
\begin{eqnarray}
(\Omega_{p})_{\sigma_{1}...\sigma_{p};\mu}=(\Omega_{p})_{\sigma_{1}...\sigma_{p},\mu}-\sum_{i}{\Gamma_{\mu\,\sigma_{i}}}^{\kappa} (\Omega_{p})_{\sigma_{1}...\sigma_{i-1}\kappa \sigma_{i+1}...\sigma_{p}}, \nonumber
\end{eqnarray}
the right hand side reads
\begin{eqnarray}
2\lambda p (\Omega_{p})_{\sigma_{1}...\sigma_{p-1}}dx_{\mu}\wedge dx^{\sigma_{1}}\wedge...\wedge dx^{\sigma_{p-1}}= 2(-1)^{(p-1)}\lambda p \,g_{\mu \sigma_{p}}\,(\Omega_{p})_{\sigma_{1}...\sigma_{p-1}}dx^{\sigma_{1}}\wedge...\wedge dx^{\sigma_{p}} \nonumber
\end{eqnarray}
and finally
\begin{eqnarray}
(\Omega_{p})_{\sigma_{1}...\sigma_{p};\mu}= 2(-1)^{(p-1)}\lambda p \,g_{\mu \sigma_{p}}\,(\Omega_{p})_{\sigma_{1}...\sigma_{p-1}}.
\end{eqnarray}
Easily (17) becomes in component notation as
\begin{eqnarray}
(\Omega_{p_*})_{\sigma_{1}...\sigma_{p_*};\mu}= 2(-1)^{p_*}\frac{\lambda}{p_*+1}\,(\Omega_{p_*+1})_{\sigma_{1}...\sigma_{p_*}\mu}.
\end{eqnarray}

\section{The derivations of equations (33) and (34)}
\textbf{Derivation of (33)}: $$\nabla_{X_a}\nabla_{X_b}\sbl {p}=2 \lambda \big(\nabla_{X_a}{e_b}\wedge \sbl {p-1}+{e_b}\wedge\nabla_{X_a}\sbl {p-1}\big)$$
$$\qquad\qquad\quad\quad\quad\quad\:\:\:\,\,=2 \lambda \big(-\omega_{bc}(X_a){e^c}\wedge \sbl {p-1}+2 \lambda {e_b}\wedge i_{X_a}\sbl {p}\big)$$
$$\qquad\qquad\quad\quad\quad\:\,=2 \lambda {\omega^{c}}_{b}(X_a)\nabla_{X_c} \sbl {p}+4 \lambda^2 {e_b}\wedge i_{X_a}\sbl {p}$$
$$\qquad\qquad\quad\quad\:\:\,= \nabla_{\nabla_{X_a}X_b} \sbl {p}+4 \lambda^2 {e_b}\wedge i_{X_a}\sbl {p}$$
so this requires
$$\big(\nabla_{X_a}\nabla_{X_b}-\nabla_{\nabla_{X_a}X_b}\big)\sbl {p}= 4 \lambda^2 {e_b}\wedge i_{X_a}\sbl {p}$$
the left hand side of which is the action of Hessian.\\

\textbf{Derivation of (34)}: The required equation can directly be deduced from the Hessian formula but here we want to follow an other way. Applying curvature operator to the degree $p$ spinor bilinear, the proof follows as: $$\mathbf{R}(X_a,X_b)\sbl {p}=((\mathbf{R}(X_a,X_b)\psi)\overline{\psi})+(\psi(\overline{\mathbf{R}(X_a,X_b)\psi}))$$
$$=-2 \lambda^2[(e_a e_b. \psi \overline{\psi})_{p}+(\psi \overline{\psi}.e_b e_a)_{p}]=-2 \lambda^2(e_a e_b. \psi \overline{\psi}+\psi \overline{\psi}.e_b e_a)_{p}$$
$$=-2 \lambda^2[e_a\wedge (e_b. \psi \overline{\psi})_{p-1}+i_{X_a}(e_b. \psi \overline{\psi})_{p+1}+e_a\wedge (\psi \overline{\psi}.e_b)^{\eta}_{p-1}-i_{X_a}(\psi \overline{\psi}.e_b)^{\eta}_{p+1}]$$
$$=-2 \lambda^2[e_{ab}\wedge (\psi \overline{\psi})_{p-2}+e_a\wedge i_{X_b} \sbl {p}+i_{X_a}(e_b \wedge (\psi \overline{\psi})_{p})+i_{X_a}i_{X_b} \sbl {p+2}$$
$$-e_{ab}\wedge (\psi \overline{\psi})_{p-2}+e_a\wedge i_{X_b} \sbl {p}+i_{X_a}(e_b \wedge (\psi \overline{\psi})_{p})-i_{X_a}i_{X_b} \sbl {p+2}],$$
after cancelations and using the fact that here $a\neq b$ then (34) is obtained.

\acknowledgments
I would like to thank \"{U}mit Ertem for helpful discussions.


\begin{thebibliography}{99}

\bibitem{Acik Ertem 2015} \"{O}. A\c{c}{\i}k and \"{U}. Ertem, ''Higher degree Dirac currents of twistor and Killing spinors in supergravity theories'', Class. Quantum Grav. \textbf{32}, 175007 (2015); \"{O}. A\c{c}{\i}k and \"{U}. Ertem, ''Generating dynamical bosons from kinematical fermions'', CQG+, (19 August 2015).

\bibitem{Acik JMP 2016} \"{O}. A\c{c}{\i}k, ''New developments in Killing spinor programme and more motivations for physics'', arXiv:1611.04424v2.

\bibitem{Baum Leitner} H. Baum, F. Leitner, ''The twistor equation in Lorentzian spin geometry'', Math. Z. 247 (2004) 795–812.

\bibitem{Lichewski} A. Lischewski, ''Towards a Classification of pseudo-Riemannian Geometries Admitting Twistor Spinors'', arXiv:1303.7246v2.

\bibitem{Kath} I. Kath, \emph{Killing spinors on pseudo-Riemannian manifolds}, Habilit., Humboldt-Universit\"{a}t zu Berlin (1999).

\bibitem{Tucker Rutherford} R W Tucker, ''Extended Particles and Exterior Calculus'', Rutherford Laboratory, Chilton-Didcot-Oxon, OX11 0QX, RL-76-022 (1976), arXiv:1610.08658v1.

\bibitem{Burton} D. A. Burton, ''A primer on exterior differential calculus'', Theoret. Appl. Mech., Vol. 30, No. 2, pp. 85-162, Belgrade 2003.

\bibitem{Penrose Rindler} R. Penrose and W. Rindler, \emph{Spinors and Space-time, Vol.2} (Cambridge Univ. Press, 1984).

\bibitem{Benn Kress spin} I. M. Benn and J. Kress, ''Differential forms relating twistors to Dirac fields'', in: Differential Geometry and its Applications, Proceedings of the 10th International Conference DGA 2007, World Scientific Publishing, Singapore, 2008, pp. 573.

\bibitem{Charlton} P. Charlton, \emph{The Geometry of Pure Spinors with Applications}, PhD thesis  (University of Newcastle 1997).

\bibitem{Acik III} \"{O}. A\c{c}{\i}k, ''Killing spinor programme: Encircling Physics'' (in preparation).

\bibitem{Benn Tucker} I. M. Benn and R. W. Tucker, \emph{An Introduction to Spinors and Geometry with Applications in
Physics} (IOP Publishing Ltd, Bristol, 1987).

\bibitem{Tucker daktilo} R. W. Tucker, ''A Clifford calculus for physical field theories'', in J. S. R. Chisholm and A. K. Common (eds.), Clifford Algebras and Their Applications in Mathematical Physics (Dordrecht: D. Reidel Publishing Company 1986).

\bibitem{Toretti} R. Toretti, \emph{Relativity and Geometry}, (Dover Publications, Inc. New York 1983).

\bibitem{Benn Kress2} I. M. Benn and J. Kress, \emph{First-order Dirac symmetry operators}, \emph{Class. Quantum Grav.} \textbf{21} (2004) 427.

\bibitem{Acik Ertem Onder Vercin1} O. Acik, U. Ertem, M. Onder and A. Vercin, ''First-order symmetries of the Dirac equation in a
curved background: a unified dynamical symmetry condition'', Class. Quantum Grav. \textbf{26} (2009) 075001.

\bibitem{Ertem 2016a} \"{U}. Ertem, ''Symmetry operators of Killing spinors and superalgebras in $AdS_5$'', J. Math. Phys. \textbf{57}, 042502 (2016).

\bibitem{Ertem 2016c} \"{U}. Ertem, ''Twistor spinors and extended conformal superalgebras'', arXiv:1605.03361.

\bibitem{Hughston Penrose Sommers and Walker} L. P. Hughston, R. Penrose, P. Sommers and M. Walker, ''On a quadratic first integral for the
charged particle orbits in the charged Kerr solution'', Commun. Math. Phys. 27 303 (1972).

\bibitem{Kastor Trachen} D. Kastor and J. Traschen, ''Conserved gravitational charges from Yano tensors'', J. High Energy Phys. JHEP08(2004)045.

\bibitem{Acik Ertem Onder Vercin2} O. Acik, U. Ertem, M. Onder and A. Vercin, \emph{Basic gravitational currents and Killing-Yano
forms}, \emph{Gen. Relativ. Gravit.} \textbf{42} (2010) 2543.

\bibitem{Acik Ertem 2016} \"{O}. A\c{c}{\i}k and \"{U}. Ertem, ''Hidden symmetries and Lie algebra structures from geometric and supergravity Killing spinors'', Class. Quantum Grav. \textbf{33}, 165002 (2016).

\bibitem{Krtous et al} P. Krtou\u{s}, D. Kubiz\u{n}\'{a}k, D. N. Page and V. P. Frolov, ''Killing–Yano tensors, rank-2 Killing
tensors, and conserved quantities in higher dimensions'', J. High Energy Phys. JHEP02(2007)004.

\bibitem{Trautmann} A. Trautmann, ''Complex structures in physics'',  arXiv:math-ph/9809022v1.

\bibitem{Cariglia et. al.} M. Cariglia, P. Krtou\u{s} and D. Kubiz\u{n}\'{a}k, ''Commuting symmetry operators of the Dirac equation, Killing-Yano and Schouten-Nijenhuis brackets'', Phys. Rev. D \textbf{84}, 024004.

\bibitem{Geroch Traschen} R. Geroch and J. Traschen, ''Strings and other distributional sources in general relativity'', Phys. Rev. D 36, 1017 (1987).

\bibitem{Stachel a} J. Stachel, ''Thickening the string I'', The string perfect dust, Phys. Rev. D \textbf{21}, 2171  (1980).

\bibitem{Stachel b} J. Stachel, Thickening the string II'', The null-string dust, Phys. Rev. D \textbf{21}, 2182  (1980).

\bibitem{Tucker mem} R. W. Tucker, ''Motion of Membranes in Spacetime'', Conference on Mathematical Relativity, 238-243, Centre for Mathematics and its Applications, Mathematical Sciences Institute, The Australian National University, Canberra AUS, (1989).

\bibitem{Tucker et al} D. Hartley, R. W. Tucker, P. A. Tuckey and T. Dray, ''Tensor Distributions on Signature-changing
Space-times'', Gen. Rel. Grav., \textbf{32}, 3 (2000).

\bibitem{Onder Tucker} M. \"{O}nder and R. W. Tucker, ''Membrane Interactions and Total Mean Curvature'', Phys. Lett. B \textbf{202}(4), 501-504 (1988).

\bibitem{Fujii Yamagishi} Y. Fujii and K. Yamagishi, ''Killing spinors on spheres and hyperbolic manifolds'', J. Math. Phys. \textbf{27} (4) 1986.

\bibitem{Fulling} S. Fulling, \emph{Aspects of Quantum Field Theory in Curved Space-Time}, London Mathematical Society Student Texts 17 (1989).

\bibitem{Dirac a} P. A. M. Dirac, ''An extensible model of the electron'', Proc. Roy. Soc. of London A, \textbf{268}, 57-67 (1962).

\bibitem{Acik 2017} \"{O}. A\c{c}{\i}k, ''Twistors from Killing Spinors alias Radiation from Pair Annihilation II: Applications'', (in preparation).

\end{thebibliography}
 \end{document}